\documentclass[a4paper]{jpconf}
\usepackage{graphicx}
\def\b{\bar}
\def\d{\partial}

\def\cD{{\cal D}}

\def\m{\mu}
\def\n{\nu}

\def\t{\tau}

\def\~{\widetilde}

\def\bY3{\bar Y_{,3}}
\def\Y3{Y_{,3}}
\def\z{\zeta}
\def\Z{{\b\zeta}}
\def\Y{{\bar Y}}
\def\cZ{{\bar Z}}
\def\`{\dot}
\def\be{\begin{equation}}
\def\ee{\end{equation}}
\def\bea{\begin{eqnarray}}
\def\eea{\end{eqnarray}}

\def\fn{\footnote}
\def\bh{black hole \ }

\def\cF{{\cal F}}

\def\mn{{\mu\nu}}

\begin{document}
\title{Fluctuating Twistor-Beam Solutions and Holographic Pre-Quantum Kerr-Schild Geometry.}

\author{Alexander Burinskii}

\address{Laboratory of Theoretical Physics, NSI Russian Academy of Sciences,\\
B.Tulskaya 52, Moscow, 115191, Russia}

\ead{bur@ibrae.ac.ru}

\begin{abstract}
Kerr-Schild (KS) geometry is based on a congruence of twistor null lines which
forms a holographic space-time determined by the Kerr theorem. We describe in details
integration of the non-stationary Debney-Kerr-Schild equations for electromagnetic
excitations of black-holes taking into account the consistent back-reaction to metric.
The exact KS solutions have the form of singular beam-like pulses supported on
twistor null lines of the Kerr congruence. These twistor-beam pulses have very strong
back reaction to metric and BH horizon and produce a fluctuating holographic KS
geometry which takes an intermediate position between the Classical and Quantum gravity.
\end{abstract}

\section{Introduction}
Singular pp-waves \cite{KraSte}, playing the role of plane waves
in gravity, differ drastically from plane waves in flat spacetime,
which is one of the origin of incompatibility of the Quantum
theory and Gravity. Quantum theory works basically in momentum
space, while Gravity demands explicit representation in
configuration space-time. Twistor theory forms a bridge between them.
Geometrically, twistor is a null line formed by a pair $(x^\m , \
\theta^\alpha),$ where $\theta^\alpha$ is a two-component spinor
joined to the point $x^\m \in M^4 .$ This spinor fixes a null
direction $\sigma^\m_{\dot\alpha \alpha} \bar\theta^{\dot\alpha}
\theta ^\alpha $ corresponding to momentum $p^\m $ of a massless
particle.
 A plane wave in twistor
 coordinates  $ T^I =\{\theta^\alpha, \ \mu_{\dot \alpha}
\}, \quad \mu_{\dot \alpha} = x_\n \sigma^\n_{\dot\alpha \alpha}
\theta^\alpha$ has the form $\exp \{i x_\m \sigma^\m_{\dot\alpha
\alpha} \tilde\theta^{\dot\alpha} \theta ^\alpha \} $ and may be
transformed to twistor space by a `twistor' Fourier transform.
 The result, \cite{Wit}(p.2.5), corresponds to a singular beam in
direction $p^\m$ supported on the twistor  null line with
coordinates $T^I =\{\theta^\alpha, \ \mu_{\dot \alpha} \}.$ Such
twistor-beams are the sources of singular pp-wave solutions in
gravity, and similar twistor-beams are obtained for typical exact
electromagnetic (em) excitations of black-holes (BH) \cite{BurA}
presented in the KS class  of the algebraically special solutions
\cite{DKS} which cover a wide range of the rotating and non-rotating
BH's and cosmological solutions.
Appearance of the twistor-beams in the exact solutions is not
accidental, since twistors form a skeleton of the KS space-time.

 In this work we give a detailed description of integration of
 the Debney-Kerr-Schild equations \cite{DKS} which describe the
 exact em excitations of the Kerr black-hole and  corresponding
 back reaction to metric\fn{In this treatment we neglect
 by the recoil of the excitations on the position of the BH.} consistent
 with the Einstein-Maxwell equations with averaged stress-energy tensor.

 The  KS geometry has twosheeted holographic twistor structure which
turns out to be perfectly adapted to quantum treatment suggested
in \cite{Gib,StW}, and we show that the resulting set of the
solutions forms a fluctuating pre-quantum geometry
taking intermediate position between the Classical and Quantum
gravity.

\section{Twistor structure of the KS geometry}

 The KS solutions are based on the KS form of metric
\be g_\mn =\eta_\mn + 2H k_\m k_\n , \label{KS}\ee in which
$\eta_\mn$ is metric of auxiliary Minkowski space-time $M^4 $ and
$k_\m$ is a field of null directions, forming a  principal null
congruence (PNC) $\cal K .$ The BH solutions may be represented in
two different forms $ g_\mn^\pm =\eta_\mn + 2H k_\m^\pm k_\n^\pm,
$ were $k^{\m\pm}(x), \ (x=x^\m \in M^4)$ correspond to two
different vector fields tangent to different congruences $\cal
K^{\pm} .$  The directions $k_\m^{\pm}$ are determined in the
Cartesian null coordinates $ \ u=(z-t)/\sqrt {2},\quad v=(z+t)/\sqrt
{2},\quad\zeta=(x+iy)/\sqrt {2},\quad\bar\zeta=(x-iy)/\sqrt {2} $
by the form \be k_\m^{(\pm)} dx^\m = P^{-1}(du +\Y^\pm d\z + Y^\pm
d\Z - Y^\pm \Y^{(\pm)} dv) \label{kpm}\ee which depends on the
complex angular coordinate $ Y=e^{i\phi} \tan \frac {\theta}{2} .$
Two solutions $Y^\pm(x)$ are determined by the {\bf Kerr Theorem}
\cite{KraSte,DKS,Pen,BurTwi} which states that the (necessary for
solutions of type D) geodesic and shear-free null congruences in
$M^4$ are generated by algebraic equation $ F = 0 , $ where
$F(Z^p)$ is an arbitrary holomorphic function of the projective
twistor coordinates.\fn{Projective twistor coordinates $ Z^p, \
p=1,2,3 $ are related with twistor coordinates $\{\theta^\alpha, \
\mu_{\dot \alpha} \}$ as follows $T^I= \theta^1 ( 1,Z^p).$}

The Kerr congruence represents a vortex of null lines, which covers
the spacetime twice: in the form of ingoing ( $k^{\m -} \in \cal
K^- $) and outgoing vector fields ($k^{\m +} \in \cal K^+ $). It forms two
sheets of the KS geometry with two different metrics $g_\mn^\pm$
on the same spacetime $M^4 .$

\begin{figure}[h]
\begin{minipage}{14pc}
\includegraphics[width=14pc]{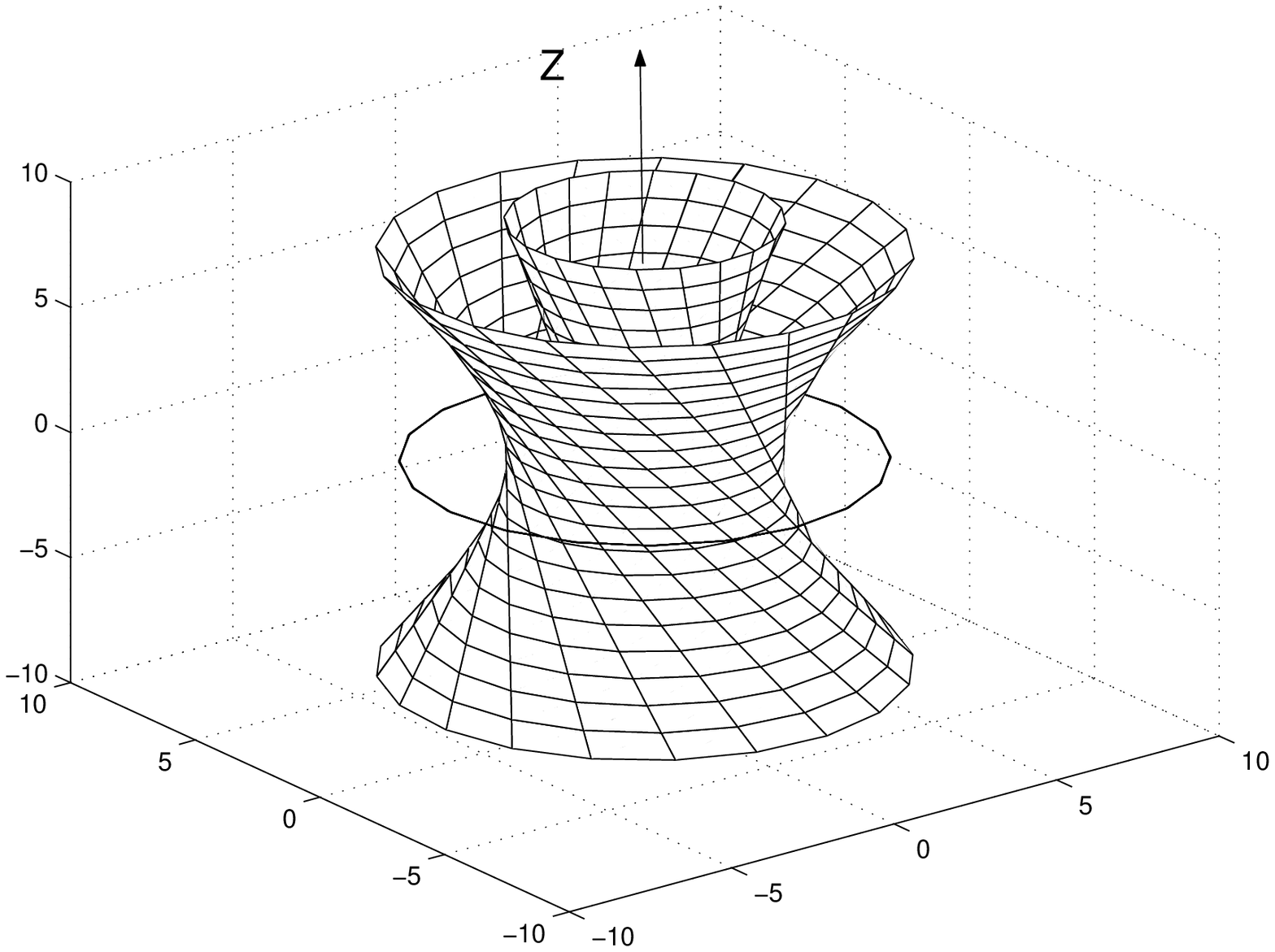}
\caption{\label{Sing}The Kerr singular ring and the Kerr
congruence of twistor null lines.}
\end{minipage}\hspace{2pc}%
\begin{minipage}{14pc}
\includegraphics[width=14pc]{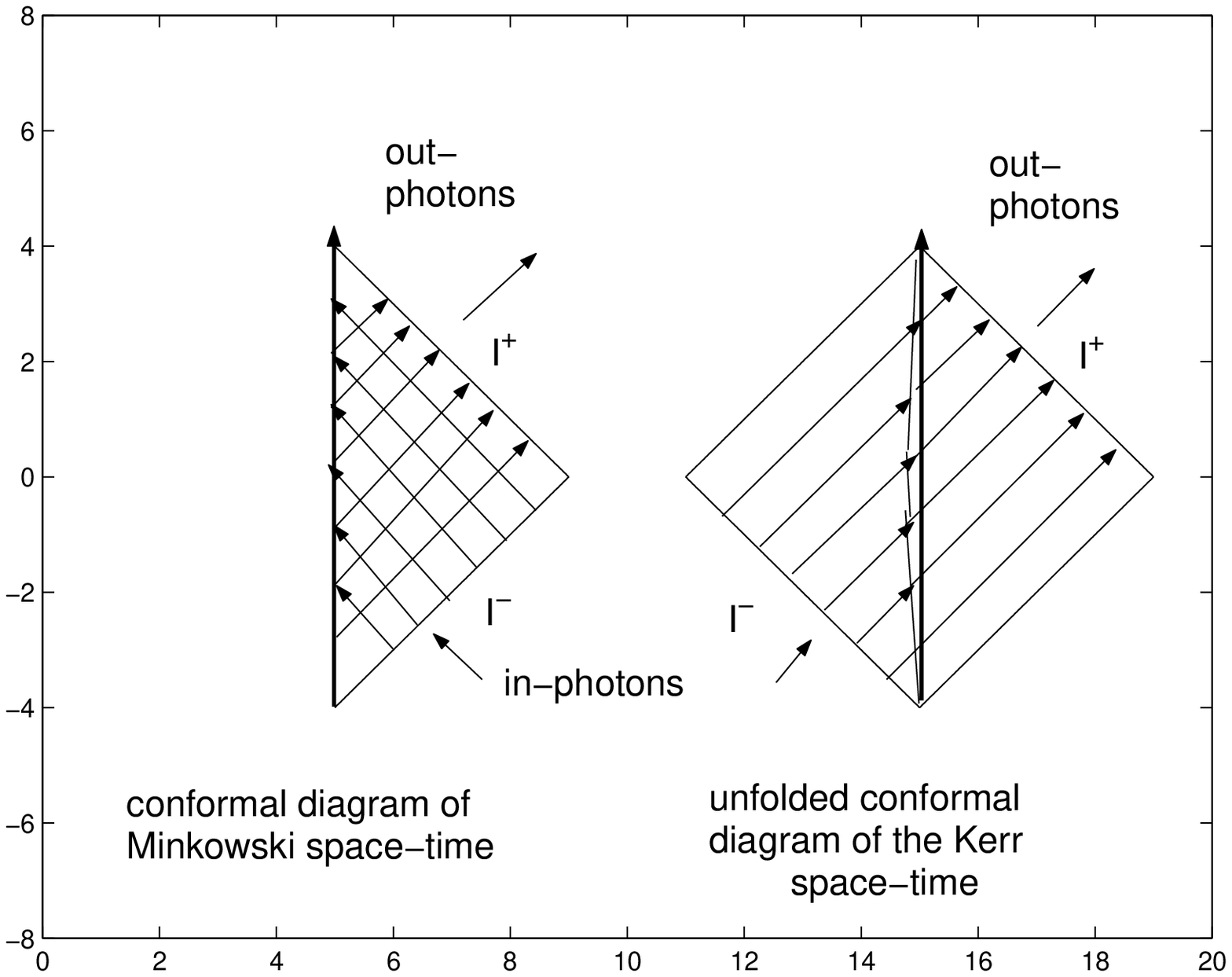}
\caption{\label{Unfold}Penrose conformal diagrams. Unfolding of
the auxiliary $M^4$ space of the Kerr spacetime yields twosheeted
structure of a pre-quantum BH spacetime.}
\end{minipage}
\end{figure}

The em field has to be aligned with the rays of PNC and turns out
to be different on the in- and out- sheets, and these two fields
should not be mixed, which is ignored usually in the
perturbative approach, leading to drastic discrepancy in the form of
the fundamental solutions. The typical  exact em solutions on the Kerr
background have the form of singular twistor-beams propagating along the
twistor null rays of the Kerr PNC, contrary to the smooth angular dependence
of the typical wave solutions used perturbatively!

 \section{Holographic KS structure}

 For a long time this twosheetedness was a mystery of the Kerr BH
(see refs in \cite{BEHM3}). It was suggested by Israel (1968) to
truncate the second sheet and replace it by a rotating disk-like bubble
covering the Kerr singular ring. Alternatively, the Kerr singular
ring was considered as a closed `Alice' string forming a gate to
`Alice' mirror world of the advanced fields \cite{BurAdv}.
Holographic approach resolves this problem unifying the both
points of view: the source of Kerr solution has to be considered
as a membrane separating the in- and out- parts of the KS space.
It is supported by the membrane paradigm and by the obtained recently
structure of the consistent Kerr-Newman source \cite{BurSuper}.
The both sheets of the KS space are necessary for description of
quantum fields in gravity. In particular, Gibbons states in
\cite{Gib} that the curved spacetime $\cal M$ should be separated
into two time-ordered regions $\cal M_-$ and $\cal M_+$ which are
associated with ingoing and outgoing vacuum states $|0_->$ and
$|0_+> .$  If the source is absent, the KS basic solutions may be
extended analytically from in- to out-sheet and vice versa.
Presence of the source breaks this analyticity, separating the
retarded and advanced fields, which allows one to consider the BH
evaporation as a scattering of the in-vacuum on the BH-source.
Similarly, in \cite{StW} authors consider a pre-quantum BH
spacetime with separated the in- and out- sheets which correspond
to a holographic correspondence: the source forms a
holographically dual boundary (membrane) separating the in- and
out- regions.\fn{In the model \cite{BurSuper} it is a domain wall
boundary of a rotating disk-like source forming a superconducting
bubble.}  The usual Penrose conformal diagram,
containing the in- and out-fields on the same of $M^4 ,$ has to be
unfolded in the holographic interpretation to split
twosheetedness, as it is shown on the Fig.2. The null rays of Kerr
congruence perform the lightlike projection of the past null
infinity $I^- ,$ so the Kerr source appears as a holographic image
of the data on the past null infinity $I^- .$

 Conformal structure of the KN solution is determined by complex function $Y(x).$
  Tetrad derivatives $\d _a = e_a^\m\d_\m $ of the function $Y(x)$
  determine principal parameters of the KS holographic projection.
  In particular, function $ Y (x)$ for the shear-free and geodesic
  congruences of the Kerr theorem has to satisfy the conditions
  \be Y,_2=Y,_4=0 , \label{Y24} \ee which
show that the expression $dY=Y,_a e^a = Y,_1 e^1 +Y,_3 e^3$
is gradient of the complex null surfaces $Y=const.$, which form a fiber bundle
of the KS space-time with fibers spanned by the tetrad forms $e^1$ and $e^3 ,$
providing conformal properties of the KS holographic projection.

\section{Stationary twistor-beam KS solutions}

\noindent For the Kerr congruence function $F(Y,x)$ of the Kerr
theorem is to be quadratic in $Y,$ and the eq. $F=0$ may resolved
in explicit form.  The corresponding solutions $Y(x)$ determine
the stationary Kerr congruence via (\ref{kpm}) and the adapted
(oblate spheroidal) coordinate system $r, \theta, \phi$ in which
\be Y=e^{i\phi} \tan \frac \theta 2 . \label{Y}\ee Function
$F(Y,x)$ determines also the function $H$ up to an arbitrary
function $\psi(Y),$ and therefore, the KS metric (\ref{KS}). In
agreement with \cite{DKS} \be H = \frac 1 2 [ m (Z + \bar Z)P^{-1}
- |\psi|^2 Z \bar Z P^{-2}] , \ee where $Z=Y,_1$ and $(Z/P)^{-1} =
- dF/dY.$ Complex function $Z=\rho + i \omega$ characterizes the
conformal properties of the congruence: expansion $\rho$ and
rotation $\omega$ of the projected image. For the Kerr-Newman BH
solution at rest, $Z$ is inversely proportional to a complex radial
distance $ Z= -P /(r+ia\cos \theta)$ where function $P$ is an
extra conformal factor determined by Killing vector of the
solution \cite{DKS}.

 In particular, for the Kerr and Kerr-Newman (KN) geometries at rest
 \be H =\frac {mr - |\psi|^2/2} {r^2+
a^2 \cos^2\theta} \ . \label{Hpsi} \ee

The KN electromagnetic field is determined by the vector potential
\be \alpha =\alpha _\m dx^\m \\
= -\frac 12 Re \ [(\frac \psi {r+ia \cos \theta}) e^3 + \chi d \Y
],   \label{alpha} \ee where $\chi = 2\int (1+Y\Y)^{-2} \psi dY  \
, $ which obeys the alignment condition $ \alpha _\m k^\m=0 . $

For the Kerr solution $m$ is mass of BH and $\psi =0 .$ For the
 KN solution  $\psi=q=const.$.
However, any nonconstant holomorphic function $\psi(Y) $ yields
also an exact KS solution, \cite{DKS}. On the other hand, any
nonconstant holomorphic functions on sphere acquire at least one
pole. A single pole at $Y=Y_i ,$  $\psi_i(Y) = q_i/(Y-Y_i)$
produces the beam in angular directions $ Y_i=e^{i\phi_i} \tan
\frac {\theta_i}{2} \label{Yi} .$ The function $\psi(Y)$ acts
immediately on the function $H$ which determines the metric and
 position of the horizon.
 The given in \cite{BEHM2} analysis showed
 that electromagnetic beams have very strong back reaction to metric
 and deform topologically the horizon, forming in the horizon the holes
 which allow matter to escape interior. For $ \psi (Y) = \sum _i \frac
{q_i} {Y-Y_i}, $ the exact solutions have several beams in angular
directions $Y_i=e^{i\phi_i} \tan \frac {\theta_i}{2},$  leading to
the horizon with many holes.
 In far zone the twistor beams tend to
the known exact singular pp-wave solutions \cite{KraSte}.

\section{ Pre-quantum fluctuating KS geometry}

The stationary KS beamlike solutions may be generalized to
time-dependent wave pulses, \cite{BurA}. Since the horizon is
extra sensitive to electromagnetic excitations, it may also be
sensitive to the vacuum electromagnetic field,  and the vacuum
beam pulses may produce a fine-grained structure of fluctuating
microholes in the horizon, allowing radiation to escape interior
of black-hole.

\begin{figure}[h]
\includegraphics[width=14pc]{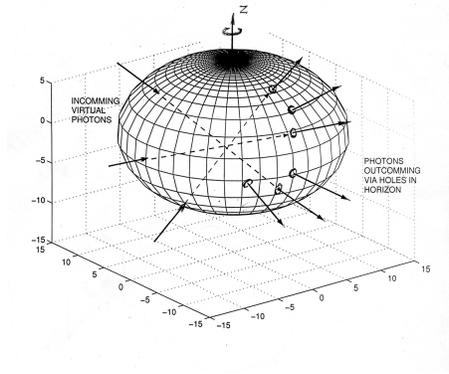}\hspace{2pc}%
\begin{minipage}[b]{14pc}\caption{\label{Vach}
Twistor-beam pulses perforate the BH horizon forming its
fluctuating fine-grained structure.}
\end{minipage}
\end{figure}

Twistor-beam pulses depend on a retarded time $\t $ and have to obey
to the non-stationary Debney-Kerr-Schild (DKS) equations \cite{DKS}.
The corresponding solutions acquire an extra radiative term
$\gamma(Y,\t) .$  The long-term attack on the DKS equations
\cite{BurA}, has led to the obtained time-dependent solutions of
the fluctuating Kerr-Schild space-times.  Derivation of
the time-dependent solutions will be given in the next section.
Here we give a brief preliminary description of the general
structure of the DKS solutions obtained by integration in \cite{DKS}.

The general em field is described by the self-dual
tetrad components, \be\cF _{12}=AZ^2, \quad \cF _{31}=\gamma Z -
(AZ),_1, \label{E6}\ee where $\cF_{ab}= e_a^\m e_b^\n \cF_\mn .$
  We will assume that the mass of BH is much greater then the
energy of the BH excitation. It allows us to neglect by recoil and
consider the  Kerr congruence as stationary. Fluctuation of the
metric will be related with fluctuation of the em field. By these
assumptions we obtain, \cite{BurA}, that the exact time-dependent
solutions describe the em radiation from BH which
contains two components:

a) a set of the singular beam pulses (determined by function
$\psi(Y,\t) $) propagating along the Kerr PNC and breaking the
topology and impenetrability of the horizon; and

b)the regularized radiative component (determined by
$\gamma_{reg}(Y,\t)$) which is smooth and determines evaporation
of the black-hole.

 The obtained solutions describe excitations of singular
electromagnetic beams $A=\psi(Y,\t)/P^2 $ on the Kerr-Schild
background and  the back reaction of the singular field
$\psi(Y,\t)$ on the metric leading to fine-grained fluctuations of the
metric and black-hole horizon.

 The most essential difference from the stationary case
is the appearance of an extra field
$\gamma (Y,\t).$ This term is absent in the expression for (\ref{Hpsi}),
and therefore, it does not have immediate action on the KS metric.
One sees also from (\ref{E6}) that it generates the null EM
radiation along the Kerr congruence which fall off asymptotically as
$\sim Z = P/(r+i a \cos \theta) \sim 1/r ,$ and so, it is the leading
term at infinity. The corresponding components of the stress-energy
tensor \be T_\mn = T_{33}e^3_\m e^3_\n \sim = \frac 12 \bar \gamma
\gamma k_\m k_\n\label{T33} \ee describe a flow of energy along the
Kerr congruence and are known as radiation of the Vaidya `shining
star' solution \cite{KraSte}. This term appears also in the Kinnersley
`photon rocket' solution \cite{KraSte}, where it is related with
acceleration of the source.

 It shows explicitly that the horizon turns out to be covered by fluctuating
 micro-holes which make it penetrable for outgoing radiation.
 In the same time, the BH radiation
is determined by the smooth field $\gamma_{reg}(Y,\t) ,$ corresponding to
the Vaidya ``shining star''  radiation.
The holographic space-time forms a fluctuating twosheeted pre-geometry
which reflects the dynamics of singular beam pulses.
This pre-geometry is classical, but  has to be still regularized to get
the usual smooth classical space-time. In this sense, it takes an
intermediate position between the classical and quantum gravity.

Note also that this structure may be generalized by using the Kerr theorem
with the higher degree in $Y$ functions $F(Y) ,$  which leads to the
multi-particle KS solutions \cite{Multiks} having the complicate networks of
twistor-beams.

\section{Time-dependent Kerr-Schild solutions} The nearest time-dependent
generalization of (\ref{Yi}) is given by the form \be \psi(Y, \t)
=\sum_i c_i(\t) (Y-Y_i)^{-1}, \label{psinst}\ee where one
assumes that an elementary beam has $c_i(\t)=q_i(\t)e^{i\omega_i
\t }$ where  $q_i(\t)$ is amplitude and $\omega_i$ is a carrier
frequency.  Similar to stationary case, the
non-stationary Kerr-Schild solutions contain the singular beams
which change the structure of black hole horizon, but the beams
may acquire the form of time-dependent pulses. Since the mass of a \bh is
much more than the energy of its excitation, we will neglect by recoil and
take the assumption that the \bh is at rest and the Kerr congruence remains
undisturbed. It corresponds to $P=2^{-1/2}(1+Y\Y)$ and $\dot P=0 .$

The obtained in \cite{DKS} DKS general equations are the equations for
functions $A$ and $\gamma $ which determine
electromagnetic field \be A,_2 - 2 Z^{-1} \cZ Y,_3 A  = 0 , \quad
A,_4 =0 ,\label{3}\ee \be \cD A+  \cZ ^{-1} \gamma ,_2 - Z^{-1} Y,_3
\gamma =0 , \label{4}\ee and the equations for function $M$ which
determine gravitational
field taking into account the back reaction caused by
electromagnetic field \be M,_2 - 3 Z^{-1} \cZ Y,_3 M  = A\bar\gamma
\cZ ,  \label{GM2}\ee \be \cD M  = \frac 12 \gamma\bar\gamma  .
\label{6}\ee The used here operator $\cD$ is \be \cD=\d _3 - Z^{-1}
Y,_3 \d_1 - \cZ ^{-1} \Y ,_3 \d_2   \ . \label{cD} \ee

 General equations for the nonstationary EM KS solutions were
obtained by by Debney, Kerr and Schild \cite{DKS} in 1968.
 However, in \cite{DKS} the equations
were fully integrated out only for the particular case $\gamma=0 $
which corresponds to stationary solutions.\fn{The degree of difficulty of
this problem is described in the excerptions from the given by
Roy Kerr in \cite{onKerr} (see Part 3 on the page 2481) review on the
history of the Kerr solution. In this review Roy calls the eqs. (\ref{E6}),
(\ref{AP2}) and (\ref{psi24}) as ``easy" ones which {\it "...
showed that the Einstein-Maxwell field depends on three functions,
$M,A,$ and $\gamma$ ... restricted by ``hard" field equations..."}
[The ``hard" equations are the EM eq. (\ref{4}) and gravitational
equation (\ref{GM2}). AB.]
{\it... "I could not solve the later unless $\gamma=0 $ so I temporarily
took this as an additional assumption and continued. This led to
the complete charged Kerr-Schild metric metrics including, of
course, charged Kerr. The null congruence for the later is the
same as the Kerr congruence but the EM field depends on an
arbitrary function of a complex variable...If it is a more
complicated function then the EM field will probably be singular.}
[This is exactly the case of stationary beam-like KS solutions. AB.]
{\it At that point I turned the
problem over to G.C. Debney ... to see if he could solve the
problem when $\gamma \ne 0$. ... In the end it became clear that
the tree of us could not solve the general problem."}}

The first non-stationary wave EM solutions on the KS background
were obtained in \cite{BurAxi} and contained singular beams along
the $\pm z-$ half-axis. There appeared the conjecture that there is no
the exact regular time-dependent solutions at all, and for very seldom
exclusions the exact solutions should have the singular beams.

The DKS equations are simplified by using the DKS relation
\be Y,_3 =- Z P_\Y/P . \label{Y3}\ee
The equation (\ref{3}) for function $A$ takes the same form
as in stationary case, \be (AP^2),_2 = 0,\ A,_4=0 ,\label{AP2} \ee
where the general solution is $A= \psi/P^2 ,$
and function $\psi$ has to obey
\be \psi,_2 = \psi,_4 = 0 \label{psi24} .\ee
To obtain the non-stationary KS solutions
 we introduce a complex
retarded-time parameter $\t $  which
satisfies the relations \be (\t),_2 =(\t),_4 = 0 \ . \label{7}
\ee It allows us to represent the equation (\ref{3}) in the form
$ (AP^2),_2=0 \ ,$ and to get
solution  which has the same form \be A= \psi(Y,\t)/P^2 \label{A} \ee
 with the exception of the extra
dependence of function $\psi$ from the retarded-time parameter $\t$.

The principal difference from the stationary case is contained in
the second electromagnetic DKS equation. Action of operator $\cD$
on the variables $Y, \bar Y $ and $ \t$ is \be \cD Y = \cD \bar Y
= 0 . \label{10}\ee For the considered here case
$P=2^{-1/2}(1+Y\bar Y)$, and $\dot P = 0 ,$ which yields \be \cD
\t =-1/P \ \label{Dt} .\ee  Using (\ref{Y3}), (\ref{10}) and
(\ref{Dt}), we reduce the equation (\ref{4}) to the form \be \dot
A = -(\gamma P),_{\bar Y} , \label{EMgamma} \ee where $\dot {( \
)} \equiv \d_{\t}$.   Integrating this equation we obtain  \bea
\gamma = - P^{-1}\int \dot A d\bar Y =
 \frac{2^{1/2}\dot \psi} {P^2 Y} +\phi (Y,\t)/P ,
\label{12}\eea where $\phi$ is an arbitrary analytic function of $Y$
and $\t.$

This solution shows that {\it any
non-stationarity in electromagnetic field ($\dot A\ne 0$) generates
an extra function $\gamma $} which, in accord with (\ref{E6}),
generates also the lightlike electromagnetic radiation along the
Kerr congruence.
Such a radiation is well-known for the Vaidya `shining star'
solution \cite{KraSte}, in which the field $A=\psi P^{-2}$ is
absent and $\gamma$ is incoherent, being related with the loss of
total mass into radiation.
The KS analog of this Vaidya relation is obtained from (\ref{6})
by substitution

\be M=m /P^3 , \label{MmP3} \ee and using (\ref{Y3}), (\ref{10}) and
(\ref{Dt}), and also the fact that $t =\frac 12 (\t + \bar \t) ,$
 \be \dot m = - \frac 12 P^4 \gamma^{(reg)} \bar \gamma^{(reg)} .
\label{G2}\ee
 It is one of two gravitational equations determining
self-consistency of the Kerr-Schild solution \cite{DKS}. The most
important consequence following from DKS equations in the
non-stationary case is the fact that the field $\gamma$ appears
inevitable, however it does not contribute to deformation of the
horizon, since it is absent in the function $H$ of (\ref{Hpsi}).
Its back reaction on the metric is smooth and circumstantial, acting
only via the slowly decreasing mass parameter $m .$ Therefore, in
the non-stationary Kerr-Schild case we obtain that the {\it
lightlike fields, determined by functions $\psi$ and $\gamma ,$
have essentially different impact on the horizon.}

Function  $\psi= \psi(\t, Y)$ obeys the equation (\ref{psi24})
which shows that the retarded time $\t$ has to satisfy the
conditions similar to (\ref{Y24}), and therefore, gradient of $\t$
is to be aligned to congruence, \be k^\m \t,_\m =0 . \ee It was
obtained in \cite{BurAxi} that the corresponding retarded-time
parameter has the form

\be \t = t -r -ia \cos \theta .\ee

Since function $\phi$ contributes only to $\gamma$ it does not
impact on the form of the horizon too. Important role of this
function is obtained from the analysis of the  gravitational
Kerr-Schild equation (\ref{GM2}) which is reduced by
using (\ref{Y3}), (\ref{10}) and (\ref{MmP3}) to the form
 \be m,_\Y = \psi\bar \gamma P . \label{G1} \ee

If we note that
$\gamma \sim \dot \psi \approx i\omega \psi ,$ we obtain that the
r.h.s. of this equation tends to zero in the low-frequency limit,
as well as the r.h.s of the equation for $\dot m .$ So, the full
solution will tend to consistent with gravity at least in the
low-frequency limit \cite{BEHM3}.
For the simplicity we consider first the case of a single
time-dependent beam generated by a single pole at $Y = Y_i .$
In this case
\be\dot \psi_i = \dot c_i(\t)/(Y-Y_i) \ee
and the solution (\ref{12}) looks singular. However, we have still the free
function $\phi ,$ form of which and parameters (time-dependence and position of its
pole) may be tuned to cancel the pole of function $\dot \psi_i ,$ and thus,
regularize the solution.
We define the analytic in $Y $ function
\be P_i = P(Y,\Y_i) =
2^{-1/2}(1+Y\Y_i)\label{Pi} \ee and set

\be \phi_i^{(tun)} (Y,\t) = - \frac{2^{1/2}  \dot c_i(\t)} {Y
(Y-Y_i) P_i }. \label{itunphi} \ee One sees that the required
analyticity of function $ \phi^{(tun)}_i (Y,\t) $ in $Y$ is
ensured. Using the equality \be (P_i - P)/(Y_i-Y) =\frac {Y}
{\sqrt{2}}  \frac {(\Y_i-\Y)} {(Y_i-Y)} ,\ee we obtain that the
regularized solution $  \gamma_i^{(reg)} = \frac{2^{1/2}\dot
\psi_i} {P^2 Y}
  +  \phi_i^{(tun)} (Y,\t)/P   \ , $
 takes the form
\be \gamma_i^{(reg)} = \frac 1{P^2} \frac {\dot c_i} {P_i}
[\frac {\Y_i-\Y}{Y_i-Y}] \label{gamreg} .\ee

The r.h.s. of the equation (\ref{G1}) for this regular solution takes
the form \be \nonumber
  \psi_i \bar\gamma_i^{(reg)}
P = -  \frac{ c_i\dot {\bar c}_i } { P \bar P_i (\Y-\Y_i)} .
\label{rhsreg} \ee

The equation (\ref{G1}) takes the form
\be m= - \int d\Y  \frac{ c_i\dot {\bar c}_i } { P \bar P_i (\Y-\Y_i)} \ee

and may now be integrated by using the Cauchy
integral formula,

$\oint \frac {f(z)dz} {z-z'} = 2\pi i f(z') ,$  and
we obtain the expression \be  m  = m_0(Y) -
2\pi i   \frac{c_i\dot {\bar c}_i } {P_i P_{ii}} \label{mci} \ , \ee
containing free function $m_0(Y),$ and also the depending on $Y$
contribution from the residue at singular point
$\Y_i  ,$ in which \be P_{ii} = \frac 1 {\sqrt 2} (1+Y_i\Y_i) \label{Pii}  \ee
is a constant. On the real slice $P$ is real, $P_i \to P_{ii} , $ and
the real part of  $m$ takes the form

\be \Re e \ m = m_0
- i \pi    \frac{c_i\dot {\bar c}_i - \dot c_i {\bar c}_i} {P_{ii}^2} , \label{rem}\ee
where $m_0$ is a real constant. Imaginary part is

\be \Im m \ m = - i \pi  \frac{c_i\dot {\bar c}_i + \dot c_i {\bar c}_i} {P_{ii}^2}
 \ .\ee

If $c_i(\t)$ is expressed
 via slowly varying amplitude $q_i(\t)$ and the carrier frequency $\omega_i ,$
$c_i(\t)=q_i(\t)e^{-i\omega_i \t } ,$ the impact of the carrier
frequency disappears and we obtain

\be \Re e \ m = m_0 - 2\pi  \omega_i \frac{ |q_i |^2 } {  P_{ii}^2 }
- i \pi    \frac{q_i\dot {\bar q}_i - \dot q_i {\bar q}_i} {P_{ii}^2}
 \ , \ee

So, the result of integration yields the time-dependence of $m$
caused by the amplitude $q_i(\t)$ and the weak and slow imaginary contribution

\be \Im m \ m = - i \pi  \frac{q_i\dot {\bar q}_i + \dot q_i {\bar q}_i} {P_{ii}^2} \ .
\label{mqi},\ee
meaning of which is not clear at this stage. We have

\be  m = m_0 - 2\pi  \omega_i \frac{ |q_i |^2 } {  P_{ii}^2 }
- 2 \pi i   \frac{q_i\dot {\bar q}_i} {P_{ii}^2}
 =  m_0 - 2 \pi i \frac{q_i} {P_{ii}^2}
 (\dot {\bar q}_i  -i\omega_i {\bar q}_i)  \ .
\ee

 Influence of the extra contributions from
$\dot q_i(\t)$ falls off as $1/T ,$ where $T$ is effective time of  action of the beam.
This contribution is negligible in the limit of the very long beams. Although these
relations may be interpreted physically as a reaction of the mass parameter to
single pulse, they cannot be considered literally, since the imaginary mass is
inconsistent with the KS class of metrics. To provide consistency with
gravitational sector we enforced to perform an averaging of the mass term over time,
which is equivalent to averaging of the stress-energy tensor.

Further, it is known \cite{KraSte} that the gravitational equation (\ref{G2}) for
radiative solutions is not equation really, but is only a definition of the loss
of mass in radiation $\dot m  =  - \frac 12 \frac{\dot c_i\dot {\bar c}_i}{|P_{i}|^2}
\ . $ On the real slice $P_i \to P_{ii},$ and in terms of the amplitudes of
beams we obtain
 \be  \dot m  = - \frac 12 \frac {\omega_i^2 { |q_i|^2 }  +  |\dot q_i|^2 }{P_{ii}^2}
= - \frac 1 {2P_{ii}^2}(\dot q_i - i\omega_i q_i)(\dot {\bar q}_i - i\omega_i \bar q_i)
. \label{q-averrad} \ee

Similar to the Vaidya `shining star' solution \cite{KraSte}, for the case of the
many stochastic pulses, we assume their frequencies as uncorrelated. There appears
a double sum over the interacting beams and  extra beating between them with
fluctuations which drop out after averaging. The surviving radiation contains only
slow fluctuations of the mass term caused by amplitudes of the self-correlated beams,
i.e. the sum of the partial solutions pulses. Therefore, the obtained solutions turn
out to be consistent with respect to the Einstein equations with averaged r.h.s.,
which for the DKS gravitational equations is equivalent to
\be m,_\Y = <P\psi\bar \gamma^{(reg)}> , \quad \dot
m = - \frac 12 <P^2 \gamma^{(reg)} \bar \gamma^{(reg)}>. \label{averG}\ee The
averaging removes the beating of the incoherent beams, however, it does not
remove the sharp back-reaction of the beams to metric and horizon,
caused by the poles in function $\psi(Y,\t)$ in agreement with
(\ref{Hpsi}).

Therefore,  the obtained solutions are exact and consistent with
the Einstein-Maxwell system of equations with averaged stress-energy tensor.
Recall also that we have used the restriction on the absence of recoil.
The case of the non-zero recoil is very hard and was considered earlier by
Kinnersley only for the non-rotating case \cite{KraSte}. It may be very
important for the problem of scattering of spinning particles in twistor
theory.

\section*{Acknowledgments} Author thanks  the Organizers and especially
Elias Vagenas for invitation to this very interesting conference and
and the RFBR (grants 07-08-00234-a, 09-02-08353-ç) for financial support.

\end{document}